\documentclass[twocolumn, superscriptaddress, amsmath,amssymb, prc aps]{revtex4-1}

\usepackage{romanbar}
\usepackage{graphicx}% Include figure files
\usepackage{dcolumn}% Align table columns on decimal point
\usepackage{bm}% bold math
\usepackage{array}
\usepackage{graphicx}  
\usepackage{siunitx}  
\usepackage{epsfig} 
\usepackage{subfigure}
\usepackage{lineno}  
\usepackage{amsmath,amssymb}
\usepackage{comment}
\usepackage{multirow}
\usepackage{dcolumn}% Align table columns on decimal point
\usepackage{bm}% bold math
\usepackage{setspace}
\usepackage{multibib}
\includecomment{printrobustnotes}
\includecomment{printallnotes}
\usepackage{xspace}
\usepackage{color}
\usepackage{hyperref} 
\begin{document}

\title{Searching for low-mass dark matter via Migdal effect in COSINE-100}
\author{G.~Adhikari}
\affiliation{Department of Physics, University of California San Diego, La Jolla, CA 92093, USA}
\author{N.~Carlin}
\affiliation{Physics Institute, University of S\~{a}o Paulo, 05508-090, S\~{a}o Paulo, Brazil}
\author{J.~J.~Choi}
\affiliation{Department of Physics and Astronomy, Seoul National University, Seoul 08826, Republic of Korea} 
\author{S.~Choi}
\affiliation{Department of Physics and Astronomy, Seoul National University, Seoul 08826, Republic of Korea} 
\author{A.~C.~Ezeribe}
\affiliation{Department of Physics and Astronomy, University of Sheffield, Sheffield S3 7RH, United Kingdom}
\author{L.~E.~Fran{\c c}a}
\affiliation{Physics Institute, University of S\~{a}o Paulo, 05508-090, S\~{a}o Paulo, Brazil}
\author{C.~Ha}
\affiliation{Department of Physics, Chung-Ang University, Seoul 06973, Republic of Korea}
\author{I.~S.~Hahn}
\affiliation{Department of Science Education, Ewha Womans University, Seoul 03760, Republic of Korea} 
\affiliation{Center for Exotic Nuclear Studies, Institute for Basic Science (IBS), Daejeon 34126, Republic of Korea}
\affiliation{IBS School, University of Science and Technology (UST), Daejeon 34113, Republic of Korea}
\author{S.~J.~Hollick}
\affiliation{Department of Physics and Wright Laboratory, Yale University, New Haven, CT 06520, USA}
\author{E.~J.~Jeon}
\affiliation{Center for Underground Physics, Institute for Basic Science (IBS), Daejeon 34126, Republic of Korea}
\author{J.~H.~Jo}
\affiliation{Department of Physics and Wright Laboratory, Yale University, New Haven, CT 06520, USA}
\author{H.~W.~Joo}
\affiliation{Department of Physics and Astronomy, Seoul National University, Seoul 08826, Republic of Korea} 
\author{W.~G.~Kang}
\affiliation{Center for Underground Physics, Institute for Basic Science (IBS), Daejeon 34126, Republic of Korea}
\author{M.~Kauer}
\affiliation{Department of Physics and Wisconsin IceCube Particle Astrophysics Center, University of Wisconsin-Madison, Madison, WI 53706, USA}
\author{H.~Kim}
\affiliation{Center for Underground Physics, Institute for Basic Science (IBS), Daejeon 34126, Republic of Korea}
\author{H.~J.~Kim}
\affiliation{Department of Physics, Kyungpook National University, Daegu 41566, Republic of Korea}
\author{J.~Kim}
\affiliation{Department of Physics, Chung-Ang University, Seoul 06973, Republic of Korea}
\author{K.~W.~Kim}
\affiliation{Center for Underground Physics, Institute for Basic Science (IBS), Daejeon 34126, Republic of Korea}
\author{S.~H.~Kim}
\affiliation{Center for Underground Physics, Institute for Basic Science (IBS), Daejeon 34126, Republic of Korea}
\author{S.~K.~Kim}
\affiliation{Department of Physics and Astronomy, Seoul National University, Seoul 08826, Republic of Korea}
\author{W.~K.~Kim}
\affiliation{IBS School, University of Science and Technology (UST), Daejeon 34113, Republic of Korea}
\affiliation{Center for Underground Physics, Institute for Basic Science (IBS), Daejeon 34126, Republic of Korea}
\author{Y.~D.~Kim}
\affiliation{Center for Underground Physics, Institute for Basic Science (IBS), Daejeon 34126, Republic of Korea}
\affiliation{Department of Physics, Sejong University, Seoul 05006, Republic of Korea}
\affiliation{IBS School, University of Science and Technology (UST), Daejeon 34113, Republic of Korea}
\author{Y.~H.~Kim}
\affiliation{Center for Underground Physics, Institute for Basic Science (IBS), Daejeon 34126, Republic of Korea}
\affiliation{Korea Research Institute of Standards and Science, Daejeon 34113, Republic of Korea}
\affiliation{IBS School, University of Science and Technology (UST), Daejeon 34113, Republic of Korea}
\author{Y.~J.~Ko}
\email{yjko@ibs.re.kr}
\affiliation{Center for Underground Physics, Institute for Basic Science (IBS), Daejeon 34126, Republic of Korea}
\author{H.~J.~Kwon}
\affiliation{IBS School, University of Science and Technology (UST), Daejeon 34113, Republic of Korea}
\affiliation{Center for Underground Physics, Institute for Basic Science (IBS), Daejeon 34126, Republic of Korea}
\author{D.~H.~Lee}
\affiliation{Department of Physics, Kyungpook National University, Daegu 41566, Republic of Korea}
\author{E.~K.~Lee}
\affiliation{Center for Underground Physics, Institute for Basic Science (IBS), Daejeon 34126, Republic of Korea}
\author{H.~Lee}
\affiliation{IBS School, University of Science and Technology (UST), Daejeon 34113, Republic of Korea}
\affiliation{Center for Underground Physics, Institute for Basic Science (IBS), Daejeon 34126, Republic of Korea}
\author{H.~S.~Lee}
\email{hyunsulee@ibs.re.kr}
\affiliation{Center for Underground Physics, Institute for Basic Science (IBS), Daejeon 34126, Republic of Korea}
\affiliation{IBS School, University of Science and Technology (UST), Daejeon 34113, Republic of Korea}
\author{H.~Y.~Lee}
\affiliation{Center for Underground Physics, Institute for Basic Science (IBS), Daejeon 34126, Republic of Korea}
\author{I.~S.~Lee}
\affiliation{Center for Underground Physics, Institute for Basic Science (IBS), Daejeon 34126, Republic of Korea}
\author{J.~Lee}
\affiliation{Center for Underground Physics, Institute for Basic Science (IBS), Daejeon 34126, Republic of Korea}
\author{J.~Y.~Lee}
\affiliation{Department of Physics, Kyungpook National University, Daegu 41566, Republic of Korea}
\author{M.~H.~Lee}
\affiliation{Center for Underground Physics, Institute for Basic Science (IBS), Daejeon 34126, Republic of Korea}
\affiliation{IBS School, University of Science and Technology (UST), Daejeon 34113, Republic of Korea}
\author{S.~H.~Lee}
\affiliation{IBS School, University of Science and Technology (UST), Daejeon 34113, Republic of Korea}
\affiliation{Center for Underground Physics, Institute for Basic Science (IBS), Daejeon 34126, Republic of Korea}
\author{S.~M.~Lee}
\affiliation{Department of Physics and Astronomy, Seoul National University, Seoul 08826, Republic of Korea} 
\author{D.~S.~Leonard}
\affiliation{Center for Underground Physics, Institute for Basic Science (IBS), Daejeon 34126, Republic of Korea}
\author{B.~B.~Manzato}
\affiliation{Physics Institute, University of S\~{a}o Paulo, 05508-090, S\~{a}o Paulo, Brazil}
\author{R.~H.~Maruyama}
\affiliation{Department of Physics and Wright Laboratory, Yale University, New Haven, CT 06520, USA}
\author{R.~J.~Neal}
\affiliation{Department of Physics and Astronomy, University of Sheffield, Sheffield S3 7RH, United Kingdom}
\author{S.~L.~Olsen}
\affiliation{Center for Underground Physics, Institute for Basic Science (IBS), Daejeon 34126, Republic of Korea}
\author{B.~J.~Park}
\affiliation{IBS School, University of Science and Technology (UST), Daejeon 34113, Republic of Korea}
\affiliation{Center for Underground Physics, Institute for Basic Science (IBS), Daejeon 34126, Republic of Korea}
\author{H.~K.~Park}
\affiliation{Department of Accelerator Science, Korea University, Sejong 30019, Republic of Korea}
\author{H.~S.~Park}
\affiliation{Korea Research Institute of Standards and Science, Daejeon 34113, Republic of Korea}
\author{K.~S.~Park}
\affiliation{Center for Underground Physics, Institute for Basic Science (IBS), Daejeon 34126, Republic of Korea}
\author{S.~D.~Park}
\affiliation{Department of Physics, Kyungpook National University, Daegu 41566, Republic of Korea}
\author{R.~L.~C.~Pitta}
\affiliation{Physics Institute, University of S\~{a}o Paulo, 05508-090, S\~{a}o Paulo, Brazil}
\author{H.~Prihtiadi}
\affiliation{Center for Underground Physics, Institute for Basic Science (IBS), Daejeon 34126, Republic of Korea}
\author{S.~J.~Ra}
\affiliation{Center for Underground Physics, Institute for Basic Science (IBS), Daejeon 34126, Republic of Korea}
\author{C.~Rott}
\affiliation{Department of Physics, Sungkyunkwan University, Suwon 16419, Republic of Korea}
\affiliation{Department of Physics and Astronomy, University of Utah, Salt Lake City, UT 84112, USA}
\author{K.~A.~Shin}
\affiliation{Center for Underground Physics, Institute for Basic Science (IBS), Daejeon 34126, Republic of Korea}
\author{A.~Scarff}
\affiliation{Department of Physics and Astronomy, University of Sheffield, Sheffield S3 7RH, United Kingdom}
\author{N.~J.~C.~Spooner}
\affiliation{Department of Physics and Astronomy, University of Sheffield, Sheffield S3 7RH, United Kingdom}
\author{W.~G.~Thompson}
\affiliation{Department of Physics and Wright Laboratory, Yale University, New Haven, CT 06520, USA}
\author{L.~Yang}
\affiliation{Department of Physics, University of California San Diego, La Jolla, CA 92093, USA}
\author{G.~H.~Yu}
\affiliation{Department of Physics, Sungkyunkwan University, Suwon 16419, Republic of Korea}
\collaboration{COSINE-100 Collaboration}
\date{\today}

\begin{abstract}
		We report on the search for weakly interacting massive particle (WIMP) dark matter candidates in the
galactic halo that interact with sodium and iodine nuclei in the COSINE-100 experiment
and produce energetic electrons that accompany recoil nuclei via the the Migdal effect. The WIMP
mass sensitivity of previous COSINE-100 searches that relied on the detection of ionization signals
produced by target nuclei recoiling from elastic WIMP-nucleus scattering was restricted to WIMP masses
above $\sim$5\,GeV/$c^2$ by the detectors' 1\,keVee energy-electron-equivalent threshold. The search
reported here looks for recoil signals enhanced by the Migdal electrons that are ejected during the
scattering process. This is particularly effective for the detection of low-mass WIMP scattering from
the crystals' sodium nuclei in which a relatively larger fraction of the WIMP's energy is transferred to
the nucleus recoil energy and the excitation of its orbital electrons. 
In this analysis, the low-mass WIMP search window of the COSINE-100 experiment is extended to WIMP mass down to 200\,MeV/$c^2$.
The low-mass WIMP sensitivity will be further improved by lowering the analysis threshold based on a multivariable analysis technique. 
We consider the influence of these improvements and recent developments in detector performance to re-evaluate sensitivities for the
future COSINE-200 experiment. 
With a 0.2\,keVee analysis threshold and high light-yield NaI(Tl) detectors
(22\,photoelectrons/keVee), the COSINE-200 experiment can explore low-mass WIMPs down to 20 MeV/$c^2$ and
probe previously unexplored regions of parameter space.
\end{abstract}

\maketitle

\section{Introduction} 
A number of astrophysical observations provide evidence that the dominant matter component of the Universe is not ordinary matter, but rather non-baryonic dark matter~\cite{Clowe:2006eq,Ade:2015xua}. Theoretically favored dark matter candidates are weakly interacting massive particles~(WIMPs)~\cite{PhysRevLett.39.165,Goodman:1984dc}. Many direct searches for WIMP dark matter in deep underground laboratories have been performed and have yet to find a signal~\cite{PhysRevLett.118.021303,Agnese:2017njq,Aprile:2017iyp,DarkSide:2018bpj,XENON:2018voc,CRESST:2019jnq,Zyla:2020zbs}. 
In light of the absence of signal in the WIMP dark matter mass range of GeV/c$^2$ to TeV/c$^2$, there is an increasing interest in low-mass dark matter particles in the sub-GeV/c$^2$ mass range~\cite{Essig:2013lka,XENON:2016jmt,XENON:2019gfn,DAMIC:2019dcn,EDELWEISS:2020fxc,SENSEI:2020dpa,SuperCDMS:2020aus,PandaX-II:2021nsg}.

In this paper, we report on low-mass dark matter searches for WIMP-nuclei interactions by looking for electron recoils induced from secondary radiation via the Migdal process~\cite{Migdal,Ibe:2017yqa} in COSINE-100 data. Data used for COSINE-100 searches has a 1\,keVee analysis threshold~\cite{Adhikari:2020xxj,COSINE-100:2021xqn}, where the unit keVee is the electron recoil-equivalent energy in kiloelectron volts. Since the total energy in the Migdal electron and the nuclear recoil is larger than the deposited energy of typical elastic nuclear recoil, our searches are extended to WIMP masses as low as 200\,MeV/c$^2$.

In the future, this search can be enhanced by lowering the analysis threshold through multivariable analysis or deep machine learning techniques, as discussed in Section~\ref{sec:sens}, where we evaluate sensitivities of the COSINE-200 experiment, which  will have lower analysis threshold, reduced internal background by controlled crystal growth~\cite{COSINE:2020egt} and improved light yield using a novel encapsulation method~\cite{Choi:2020qcj}.

\section{Experiment} 

The COSINE-100 experiment~\cite{Adhikari:2017esn} is installed in the Yangyang underground laboratory (Y2L) utilizing the space provided by the Yangyang pumped storage power plant in South Korea~\cite{Kims:2005dol,Kim:2012rza}. The laboratory is located at a vertical depth of 700\,m that provides a water-equivalent overburden of 1800\,m~\cite{Prihtiadi:2017inr}. A 2-km-long driveway provides access to the laboratory as well as air ventilation. The laboratory is equipped with a cleanroom and an air-conditioning system providing a low dust level, and constant temperature and humidity of 24.21$\pm$0.06$^\circ$C and 36.7$\pm$1.0\%, respectively~\cite{Adhikari:2017esn,Kim:2021rsj}. 
The contamination level of $^{222}$Rn in the room is measured to 36.7$\pm$5.5\,Bq/m$^3$. The readout electronics, high voltages that are applied to photomultiplier tubes~(PMTs), and data acquisition system are also monitored and stably maintained~\cite{Kim:2021rsj}.

\begin{figure}[!htb]
\begin{center}
\includegraphics[width=0.49\textwidth]{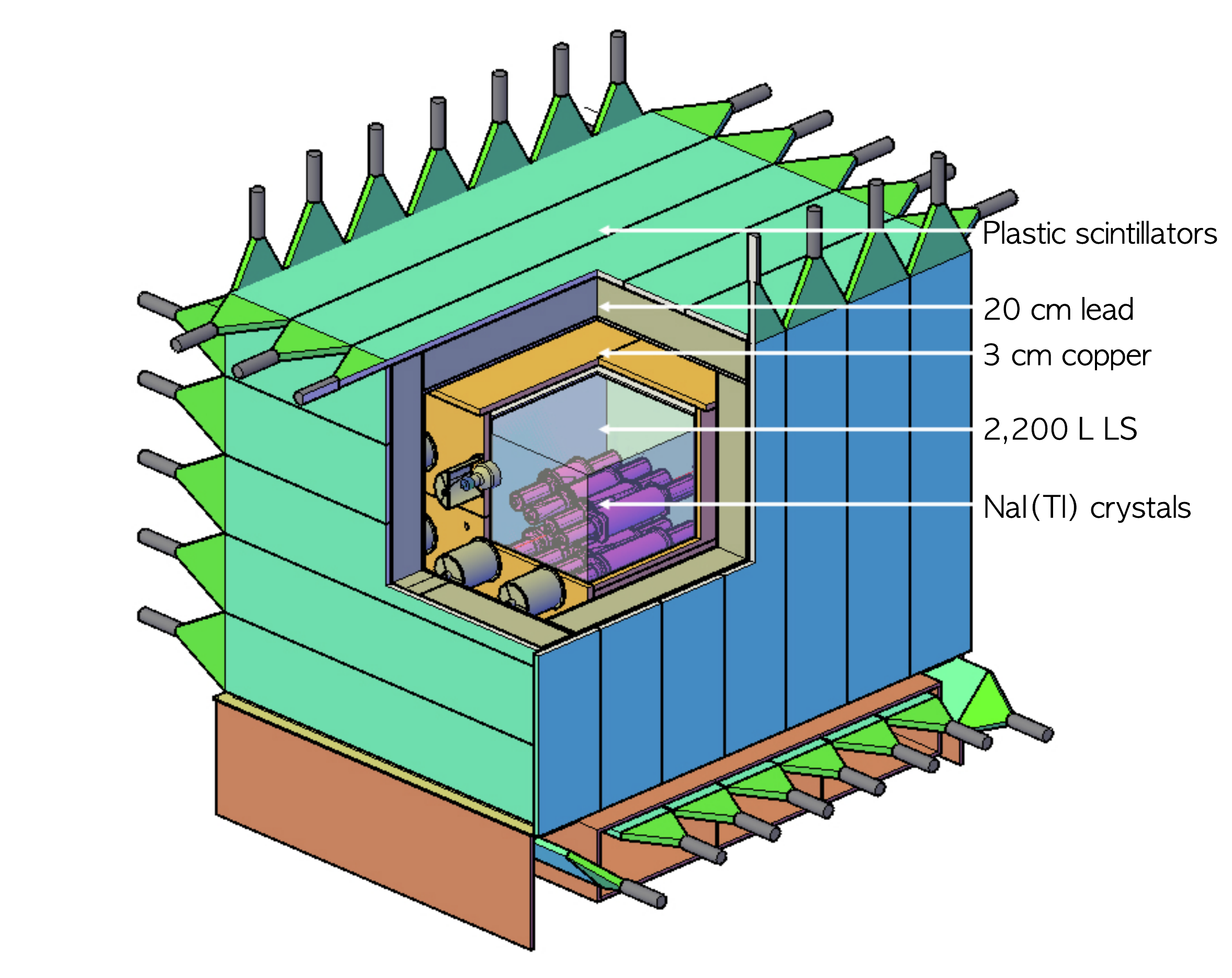} 
\caption{Schematic of the COSINE-100 detector. The NaI(Tl)~(106~kg) detectors are immersed in the 2,200\,L LAB-LS that is surrounded by layers of copper and lead shielding.}
\label{fig_det}
\end{center}
\end{figure}

The COSINE-100 detector, shown in Fig.~\ref{fig_det}, consists of a 106\,kg array of eight ultra-pure NaI(Tl) crystals each coupled to two PMTs. The crystal array is immersed in an active veto detector comprised of 2,200\,L of linear alkylbenzene~(LAB)-based liquid scintillator~(LS) to attenuate or tag the influence of external or internal radiations~\cite{Park:2017jvs,Adhikari:2020asl}. The LAB-LS is contained within a box made with 1\,cm thick acrylic and 3\,cm thick oxygen-free copper that is surrounded by a 20\,cm thick lead shield. An outer array of plastic scintillation counters is used to tag and veto cosmic-ray muons~\cite{Prihtiadi:2017inr,Prihtiadi:2020yhz}. 

An event is triggered when coincident single photoelectrons in both PMTs that are coupled to a single crystal are observed within a 200\,ns time window. If at least one crystal satisfies the trigger condition, data from all crystals and the LAB-LS are recorded. The signals from the crystal PMTs are processed by 500\,MHz flash analog-to-digital converters and are 8\,$\mu$s long waveforms that start 2.4\,$\mu$s before the trigger. The LAB-LS and plastic scintillator signals are processed by charge-sensitive flash analog-to-digital converters. Muon events are triggered by coincident signals from at least two plastic scintillators. The LAB-LS signals do not generate triggers, except in the case of energetic muon events that are coincident with one of the muon detector panels. A trigger and clock board read the trigger information from individual boards and generate a global trigger and time synchronizations for all of the modules. Details of the COSINE-100 data acquisition system are described in Ref.~\cite{Adhikari:2018fpo}. 

\section{Migdal effect}

Direct detection of WIMP dark matter with mass below sub-GeV/c$^2$ is limited by an energy threshold of the detector in the range about 0.1--1\,keVee. Because the nuclear recoil energies from WIMP-nuclei interactions are quenched (the scintillation signals from nuclear recoils are only an order of 10\,\% of the signals from the same energy deposition of electrons~\cite{Manzur:2009hp,Lee:2015xla,Joo:2018hom,Kimura:2019rdg}), this energy range corresponds to 1--10\,keVnr, where keVnr is kiloelectron volt nuclear recoil energy. 
The WIMP-nucleus interaction used in typical direct detection searches assumes that the electron cloud is tightly bound to the nucleus and that the orbit electrons remain in stable states. However, energy transferred to nuclei after collision may lead to excitation or ionization of atomic electrons via the Migdal process~\cite{Migdal,Ibe:2017yqa}. 
This process can lead to the production of energetic electron-induced signals that are produced in association with the primary nuclear recoil. 
For a WIMP-nucleus interaction, even if the electron equivalent energy implied by the quenching factor is below the energy threshold of the detector, these Midgal-effect secondary electrons can produce electron equivalent energy that are above the threshod, making detectors sensitive to sub-GeV/c$^{2}$ dark matter interactions. 
Several experimental groups have already exploited this effect to search for dark matter with sub GeV/c$^2$ masses~\cite{LUX:2018akb,EDELWEISS:2019vjv,CDEX:2019hzn,XENON:2019zpr,GrillidiCortona:2020owp}.

The differential nuclear recoil rate per unit target mass for elastic scattering between WIMPs of mass $m_\chi$ and target nuclei of mass $M$ is~\cite{Savage:2008er},
\begin{eqnarray}
		\frac{dR_\mathrm{nr}}{dE_\mathrm{nr}} = \frac{\rho_\chi}{2m_\chi\mu^2}~\sigma(M,~E_\mathrm{nr})
  \int_{v>v_\mathrm{min}}d^3v~f(\textbf{v},~t),
\label{eq:recoilrate}
\end{eqnarray}
where $\rho_\chi$ is the local dark matter density, $E_\mathrm{nr}$ is the nuclear recoil energy, $\sigma(M,E_\mathrm{nr})$ is the WIMP-nucleus cross section and $f(\textbf{v},t)$ is the time-dependent WIMP velocity distribution. The reduced mass $\mu$ is defined as $m_\chi M/(m_\chi+M)$ and the minimum WIMP velocity $v_\mathrm{min}$ is $\sqrt{ME_\mathrm{nr}/2\mu^2}$.

\begin{figure}[!htb]
\begin{center}
\includegraphics[width=0.49\textwidth]{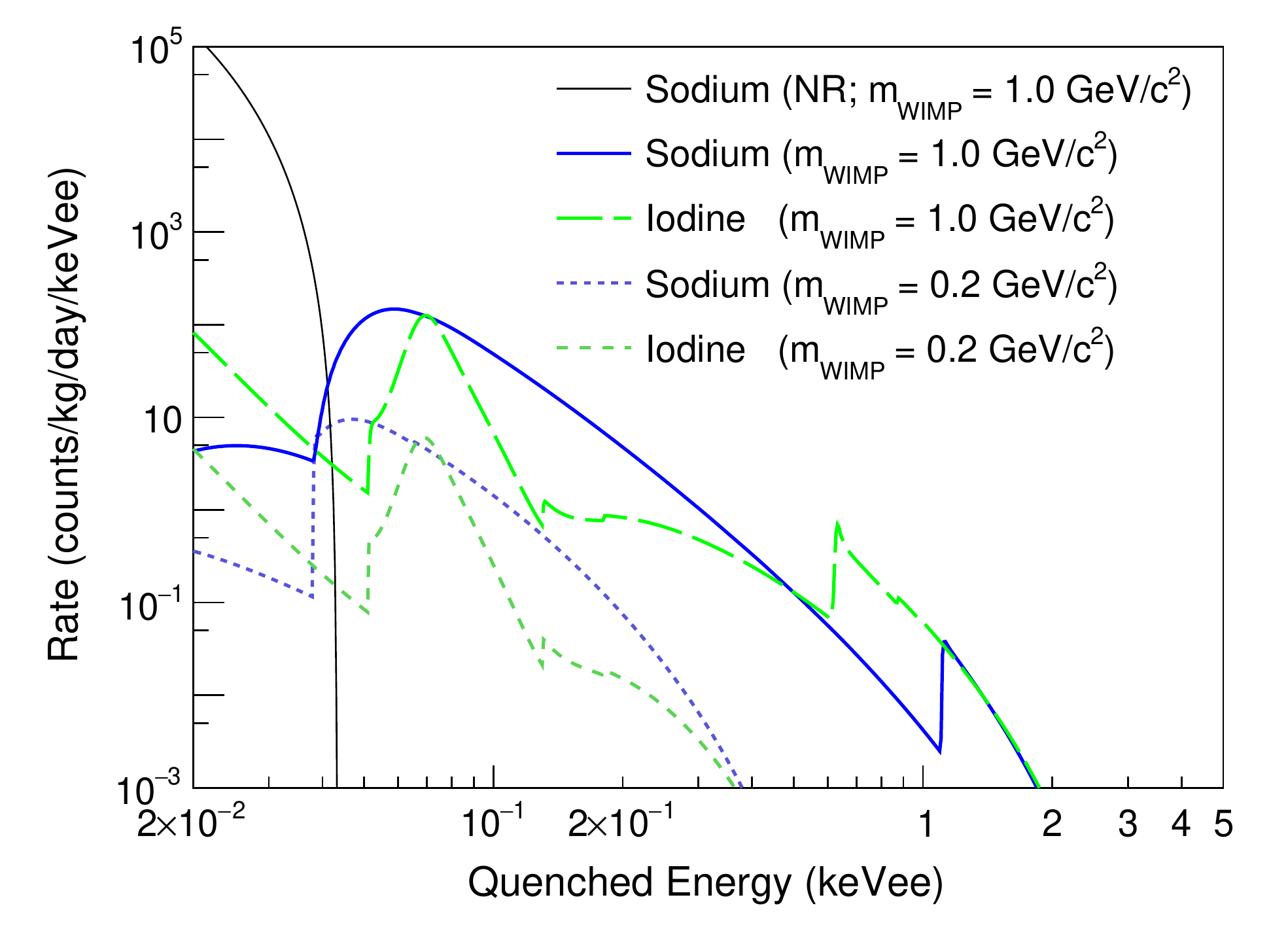}
\caption{Expected signals without energy resolution from WIMP-nuclei spin-independent (SI) interaction via the Migdal process are presented for WIMP masses of 0.2\,GeV/c$^2$ and 1.0\,GeV/c$^2$ assuming WIMP-nucleon SI cross-section of 1\,pb. Sodium and iodine spectra produced by the Migdal effect are separately presented and compared with the nuclear recoil spectrum of WIMP-sodium interaction for 1.0\,GeV/c$^2$ WIMP mass. Here the nuclear recoil energy is quenched to the electron recoil energy using the measured quenching factor reported in Ref.~\cite{Joo:2018hom}. Although the nuclear recoil deposit is below the 1\,keVee energy threshold, the electron energy from the Migdal effect can produce above-threshold signals. }
\label{fig_signal}
\end{center}
\end{figure}

The rate of ionization due to the Migdal effect for a nuclear recoil energy $E_\mathrm{nr}$ accompanied by an ionization electron with energy $E_{ee}$ is given by the nuclear recoil in Eq.~\ref{eq:recoilrate} multiplied by the ionization rate~\cite{Ibe:2017yqa},
\begin{equation}
	\begin{split}
	\frac{dR}{dE_\mathrm{ee}} & =  \int dE_\mathrm{nr} dv \frac{d^2R}{dE_\mathrm{nr}dv} \\*
	& \times \frac{1}{2\pi}\sum_{n,l} \frac{d}{dE_\mathrm{nr}} p^{c}_{q_e}(n,l \rightarrow E_\mathrm{ee}-E_{n,l}).
	\end{split}
\label{eq:migdal}
\end{equation}
Here $p^{c}_{q_e}$ is the probability for an atomic electron with quantum number ($n,l$) and binding energy $E_{n,l}$ to be ejected with a kinetic energy $E_\mathrm{ee}-E_{n,l}$, and $q_e$ is the electron momentum  in the nucleus rest frame. When the shell vacancy is refilled, an X-ray or an auger electron with energy $E_{n,l}$ is emitted. This takes into account the fact that the emitted electron may come from an inner orbital and the remaining excited state will release additional energy when returns to its ground state. The differential probability rates for sodium and iodine were calculated in Ref.~\cite{Ibe:2017yqa}. Figure~\ref{fig_signal} shows the differential ionization rates as a function of the electron recoil energy $E_\mathrm{ee}$  for sodium and iodine nuclei considering two different WIMP masses.

This description assumes isolated atomic targets that interact with WIMP particles~\cite{Ibe:2017yqa,Kouvaris:2016afs}, which for inner-shell electrons provided a correct estimate of the expected signal rate. For outer electrons, the complicated electronic band structure and crystal form factor can affect the rate for the Migdal effect and significantly improved sensitivity in semiconductors has been reported in Refs.~\cite{Essig:2019xkx,Knapen:2020aky}. 
In this analysis, we follow the atomic target approximation because at our analysis threshold, inner shell electrons are the dominant contributors to the Migdal process.

\section{Data analysis}

We use data obtained from October 2016 to July 2018, corresponding to 1.7 years exposure that were used for our first annual modulation search~\cite{Adhikari:2019off} and the model-dependent WIMP dark matter search using the energy spectra~\cite{COSINE-100:2021xqn}. During the 1.7\,years data-taking period, no significant environmental anomalies or unstable detector performance were observed. Six of the eight crystals have a high light yield of approximately 15\,NPE/keVee, where NPE corresponds to the number of photoelectron, and enables an analysis threshold of 1\,keVee. The other two crystals had lower light yields and required higher analysis thresholds~\cite{Adhikari:2017esn,Adhikari:2018ljm}. Since their direct impact on the low-energy signal search is not substantial, we do not include these two crystals in this analysis.

In the offline analysis, muon-induced events are rejected when the crystal hit events and muon candidate events in the muon detector~\cite{Prihtiadi:2017inr,Prihtiadi:2020yhz} are coincident within 30\,ms. Additionally, we require that the leading edges of the trigger pulses start later than 2.0\,$\mu$s after the start of the recording, that the waveforms from the hit crystal contain more than two single photoelectrons and the integral waveform area below the baseline does not exceed a limit. These criteria reject muon-induced phosphor events and electronic interference events. A multiple-hit event is one in which more than one crystal has a signal with more than four photoelectrons in an 8\,$\mu$s time window or, has an LS signal above an 80\,keVee threshold within 4\,$\mu$s of the hit crystal~\cite{Adhikari:2020asl}. A single-hit event is classified as one where only one crystal is hit and none the other detectors meet the above criteria.

\begin{figure*}[!htb]
\begin{center}
\includegraphics[width=0.98\textwidth]{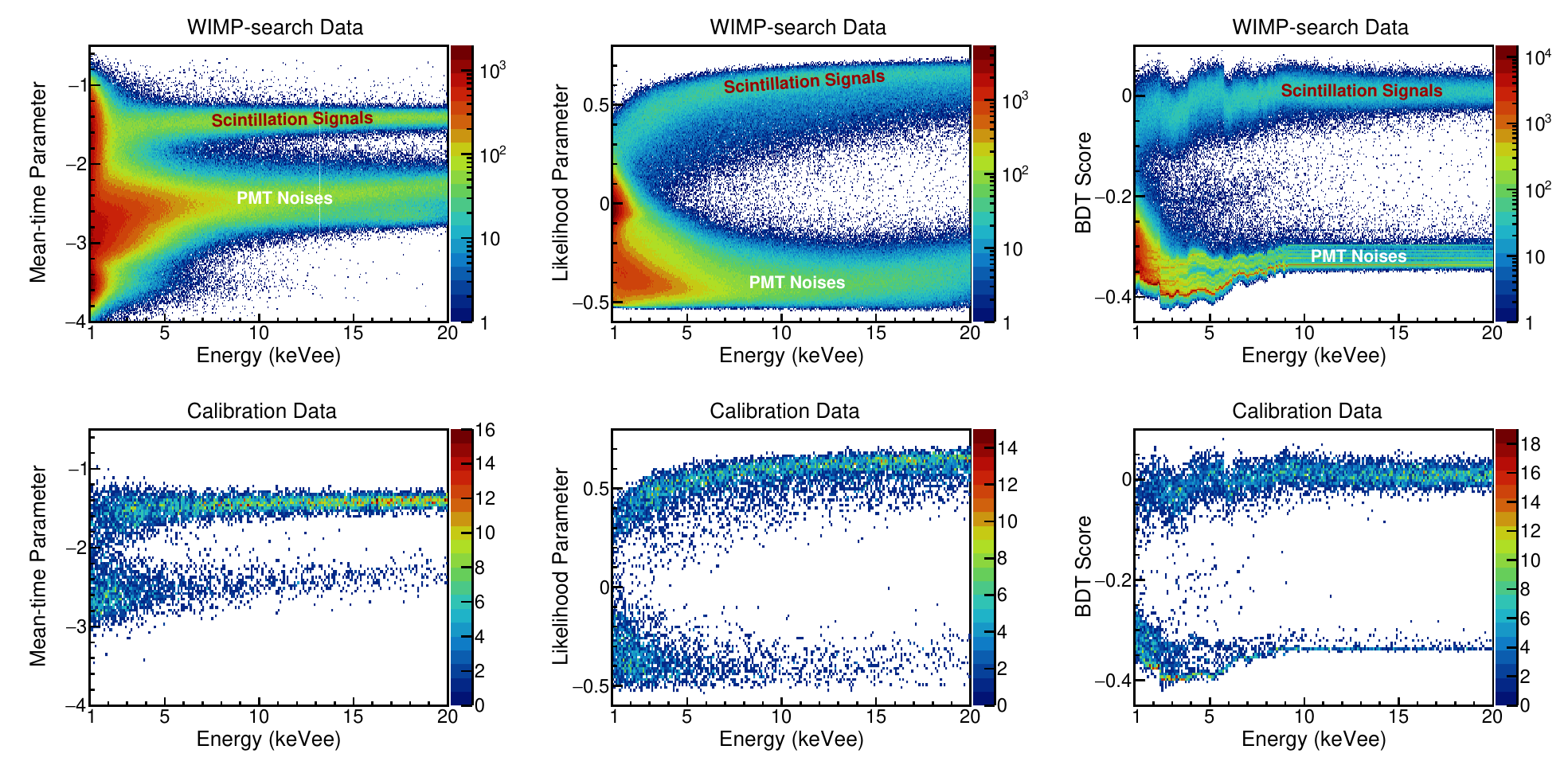} 
\caption{The distribution of two input parameters, mean-time parameter (left) and the likelihood parameter (center), and output BDT scores (right) are presented separately for the  WIMP-search data (top) and $^{60}$Co calibration data (bottom). Details of all parameters are described in~\cite{Adhikari:2020xxj}.}
\label{fig_bdtpar}
\end{center}
\end{figure*}

In the low-energy signal region below 10\,keVee, PMT-induced noise events contribute to the single-hit WIMP-search data.
Although PMT noise involves complex phenomena that are far from being completely understood, we categorize two distinct classes of the PMT-induced noise events. 
The first class has a fast decay time of less than 50\,ns, compared with typical NaI(Tl) scintillation of about 250\,ns. This class of noise events may be induced by radioactive decays of U, Th, and K inside the PMT materials. These decays may generate ultraviolet and/or visible photons inside the PMTs. The second class, that occurs less often than the first, is characterized by slow rise times~(about 100\,ns) and decay times~(about 150\,ns), as described in Refs.~\cite{Adhikari:2018ljm,Kang:2019fvz}. This class of noise may be caused by accumulated charge somewhere in the PMT arising from the high voltage and subsequent-discharging that produces a flash inside the PMTs. This class noise events are intermittently produced by certain PMTs. We have developed monitoring tools for data quality verification, including monitoring event rates of the second class of noise. If a crystal has an increased rate due to the second class of noise, the relevant period of data is removed. One crystal detector has this class of noise for the whole data-taking period; for the other five detectors the second-class noise-induced events are absent during more than 95\,\% of the data taking period. The effective data exposure for the five crystals is 97.7\,kg$\cdot$year. 

The first class of PMT noise-induced events is efficiently rejected by a boosted decision tree (BDT)-based multivariable analysis techinique~\cite{BDT}. The parameters used in the BDT include the balance of the deposited charge from two PMTs, the ratio of the leading-edge (0--50\,ns) to trailing-edge (100\,ns--600\,ns) charge, a mean-time parameter, which is a logarithm of the amplitude weighted average time of the events, and a likelihood parameter for samples of scintillation-signal events and the fast PMT-induced events~\cite{Adhikari:2020xxj,COSINE-100:2021xqn}. Figure~\ref{fig_bdtpar} shows input parameters of mean-time and likelihood, as well as output BDT scores for calibration samples and WIMP search data. 
This procedure reduces the noise contamination level to less than 0.5\% and maintains an 80\% selection efficiency at the lowest energy bin~(1--1.25\,keVee) can be achieved~\cite{Adhikari:2020xxj}. 

Geant4~\cite{Agostinelli:2002hh}-based simulations are used to understand the contribution of each background component~\cite{Adhikari:2017gbj,cosinebg,cosinebg2}, as well as to verify the energy scales and resolutions. We classify four categories of the NaI-deposited events based on their energies and detector multiplicities. Single-hit and multiple-hit events are further divided into low-energy events 1--70\,keVee and high-energy events 70--3000\,keVee. The fraction of each background component is determined from a simultaneous fit to the four categorized distributions. For the single-hit events, only 6--3000\,keVee events are used to avoid a bias of the signal in the region of interest (ROI). A detailed  description of our modeling of the background with the same data is described in Ref.~\cite{cosinebg2}.  

We consider various sources of systematic uncertainties in the background and signal models. Errors associated with the selection efficiency, the energy resolution, the energy scale, and the background modeling technique are accounted for the shapes of the signal and background probability density functions, as well as in rate changes as described in Ref.~\cite{COSINE-100:2021xqn}. These quantities are allowed to vary within their uncertainties as nuisance parameters in the data fit used to extract the signal. 

To estimate the dark matter signal enhancement that is provided by the Migdal effect, we generate signals based on Eq.~\ref{eq:migdal} for various interaction models and different masses in the specific context of the standard WIMP galactic halo model~\cite{Lewin:1995rx,Freese:2012xd}. Both a spin-independent (SI) interaction between WIMP and nucleons and a spin-dependent (SD) interaction between WIMP and proton are considered in this analysis. Because both sodium and iodine have non-zero proton spin due to their odd numbers of protons~\cite{Bednyakov:2006ux,Gresham:2014vja}, NaI(Tl) detectors are sensitive to the WIMP-proton SD interactions. Responses that include form factors and proton spin values of the nuclei are implemented from the publicly available {\sc dmdd} package~\cite{dmdd,Gluscevic:2015sqa,Anand:2013yka,Fitzpatrick:2012ix,Gresham:2014vja}. The energy spectra of electron-equivalent energy for the detector are simulated with the energy resolutions of the detectors and the nuclear recoil quenching factors (QFs), where QF is the ratio of the scintillation light yield from sodium or iodine recoil relative to that from electron recoil for the same energy. We used the QF values from recent measurements with monoenergetic neutron beam~\cite{Joo:2018hom}. The measurements were modeled using a modified Lindhard formula~\cite{osti_4701226} and described in Ref.~\cite{Ko:2019enb}. Examples of quenched signal spectra from the Migdal effect of the WIMP-nucleon SI interactions for the WIMP masses of 0.2 and 1.0\,GeV/c$^2$ are presented in Fig.~\ref{fig_signal}. The output events are subjected to the same selection criteria that are applied to the data. 

\begin{figure}[!htb]
\begin{center}
\includegraphics[width=0.49\textwidth]{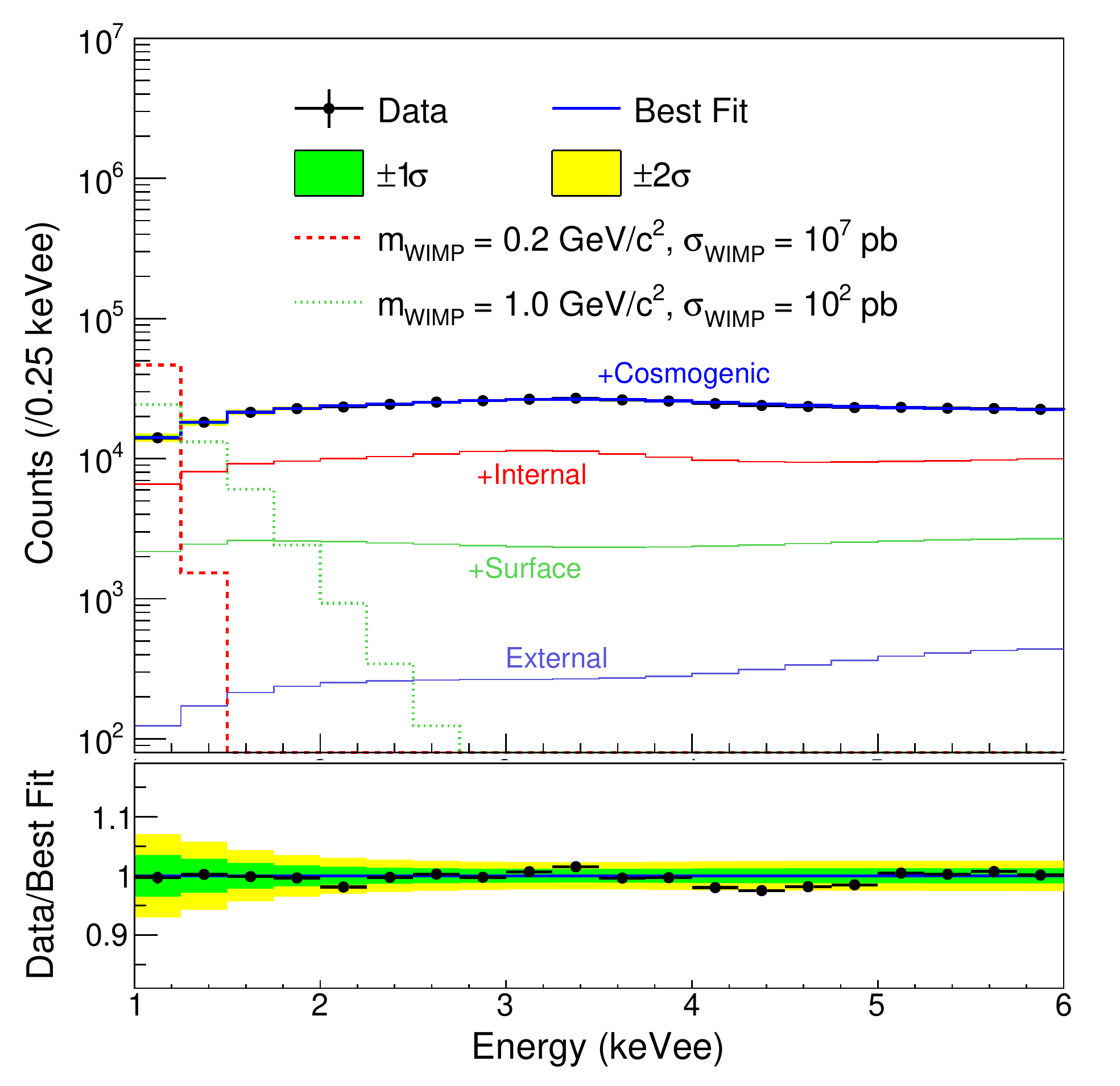}
\caption{An example of the fit for a WIMP mass 1.1\,GeV/c$^2$ using the Migdal effect is presented. The summed energy spectrum for the five crystals (black filled circles) and the best fit (blue solid line) for which no WIMP signals are obtained, are shown together with signal spectra from WIMP masses and SI cross-sections of 0.2\,GeV/c$^2$ and 10$^{7}$pb~(red dashed line), 1.0\,GeV/c$^2$ and 10$^{2}$~\,pb~(green dotted line). The fitted distribution is broken down into cumulative contributions to the background from external sources, the surface of the crystals and nearby materials, internal radionuclide contaminations, and cosmogenic activation, as indicated. The green (yellow) bands are the 68\% (95\%) CL intervals of the systematic uncertainty obtained from the likelihood fit.}
\label{fig_fit}
\end{center}
\end{figure}

\begin{figure*}[!htb]
\begin{center}
\includegraphics[width=0.98\textwidth]{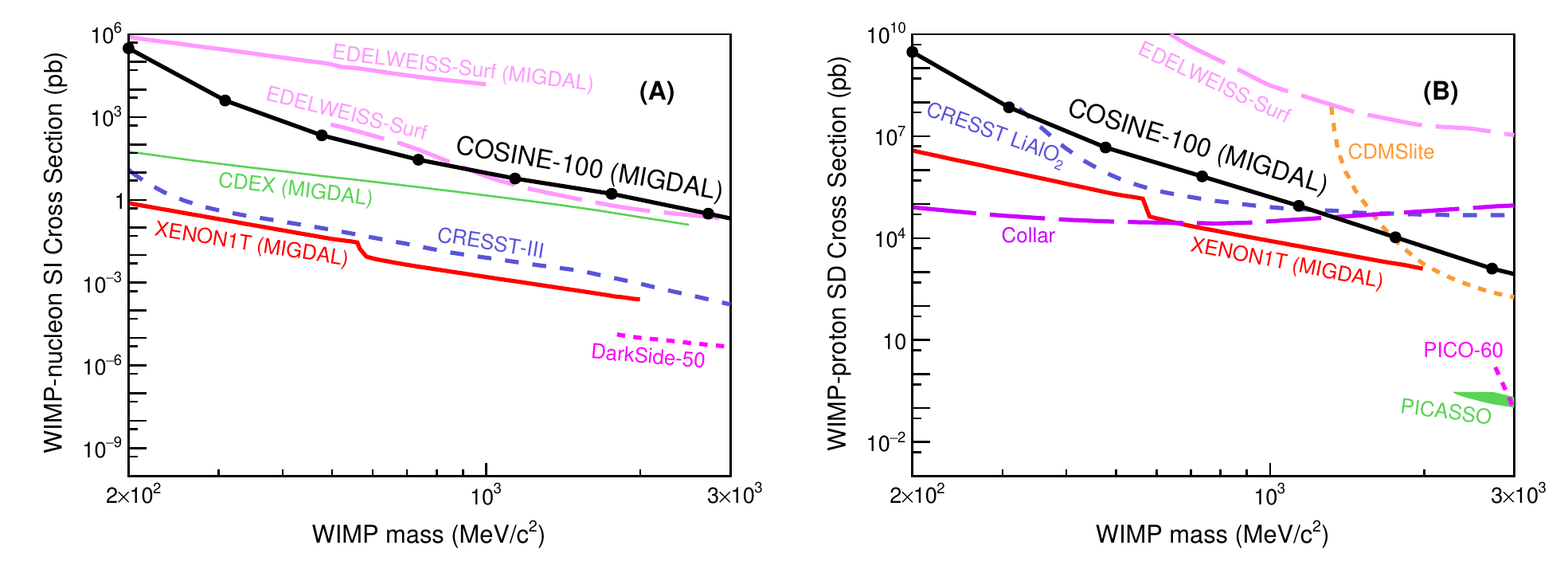} 
\caption{Limits on WIMP-nucleon SI interaction (A) and WIMP-proton SD interaction (B) induced by the Migdal effect from the COSINE-100 data (black dots) at 90\% CL are compared with XENON1T~\cite{XENON:2019zpr}, EDELWEISS surface measurement~\cite{EDELWEISS:2019vjv},  and CDEX~\cite{CDEX:2019hzn} with the Migdal effect, and CRESST-III~\cite{CRESST:2019jnq},  CRESST LiAlO$_2$G~\cite{CRESST:2020tlq}, DarkSide-50~\cite{Agnes:2018ves}, PICASSO~\cite{Behnke:2016lsk}, CDMSlite~\cite{SuperCDMS:2017nns}, EDELWEISS surface~\cite{EDELWEISS:2019vjv}, Collar~\cite{Collar:2018ydf}, and PICO-60~\cite{PICO:2019vsc}.}
\label{fig_limit}
\end{center}
\end{figure*}

A Bayesian approach is adopted to extract the WIMP interaction signals using the Migdal effect from the COSINE-100 data. A likelihood function based on Poisson probability is built including constraints that reflect the known levels of the background components. The likelihood fit is applied to the measured single-hit energy spectra between 1 and 6\,keVee for each WIMP model with various masses. Each crystal is fitted with a crystal-specific background model and a crystal-correlated dark matter signal.  The combined fit is constructed by multiplying likelihoods of the five crystals. The systematic uncertainties are included as nuisance parameters with Gaussian constraints. The same machinery was used for the WIMP dark matter searches using the same dataset but with only the energy from the nuclear recoils without the Migdal effect~\cite{COSINE-100:2021xqn}. An example of the fit for the SI interaction for a  WIMP mass of 1.1\,GeV/c$^{2}$ is shown in Fig.~\ref{fig_fit}. The summed event spectrum for the five crystals is shown together with the best-fit result. For comparison, the expected signals for WIMP masses (cross-sections) of 0.2 (10$^7$) and 1.0\,GeV/c$^2$ (10$^2$\,pb) are presented together. No significant signal for event excess that could be attributed to WIMP interactions is found and 
90\% confidence level (CL) limits are determined from the marginalization of the likelihood function. Figure~\ref{fig_limit} shows the 90\% CL upper limits from the COSINE-100 data for the WIMP-nucleon SI interactions (A) and the WIMP-proton SD interactions (B) compared with the limits from XENON1T~\cite{XENON:2019zpr}, EDELWEISS surface measurement~\cite{EDELWEISS:2019vjv}, and CDEX~\cite{CDEX:2019hzn} using the Migdal effect, as well as CRESST-III~\cite{CRESST:2019jnq}, CRESST LiAlO$_2$G~\cite{CRESST:2020tlq}, DarkSide-50~\cite{Agnes:2018ves}, PICASSO~\cite{Behnke:2016lsk}, CDMSlite~\cite{SuperCDMS:2017nns}, EDELWEISS surface~\cite{EDELWEISS:2019vjv}, Collar~\cite{Collar:2018ydf}, and PICO-60~\cite{PICO:2019vsc} results. 
Here, our search extends to the low-mass WIMP of 200\,MeV/c$^2$. 
Although our limits do not consider earth shielding effects~\cite{Kouvaris:2015laa,Kavanagh:2017cru,EDELWEISS:2019vjv}, we found no significant impact on the upper limits using the {\sc verne} code~\cite{verne}.
Because of the relatively higher threshold energy and background rates, our search cannot investigate the unexplored parameter space. This will be addressed in the future with improved detectors and analysis methods. 

\section{Sensitivity for COSINE-200}
\label{sec:sens}
\begin{figure}[!htb]
\begin{center}
\includegraphics[width=0.49\textwidth]{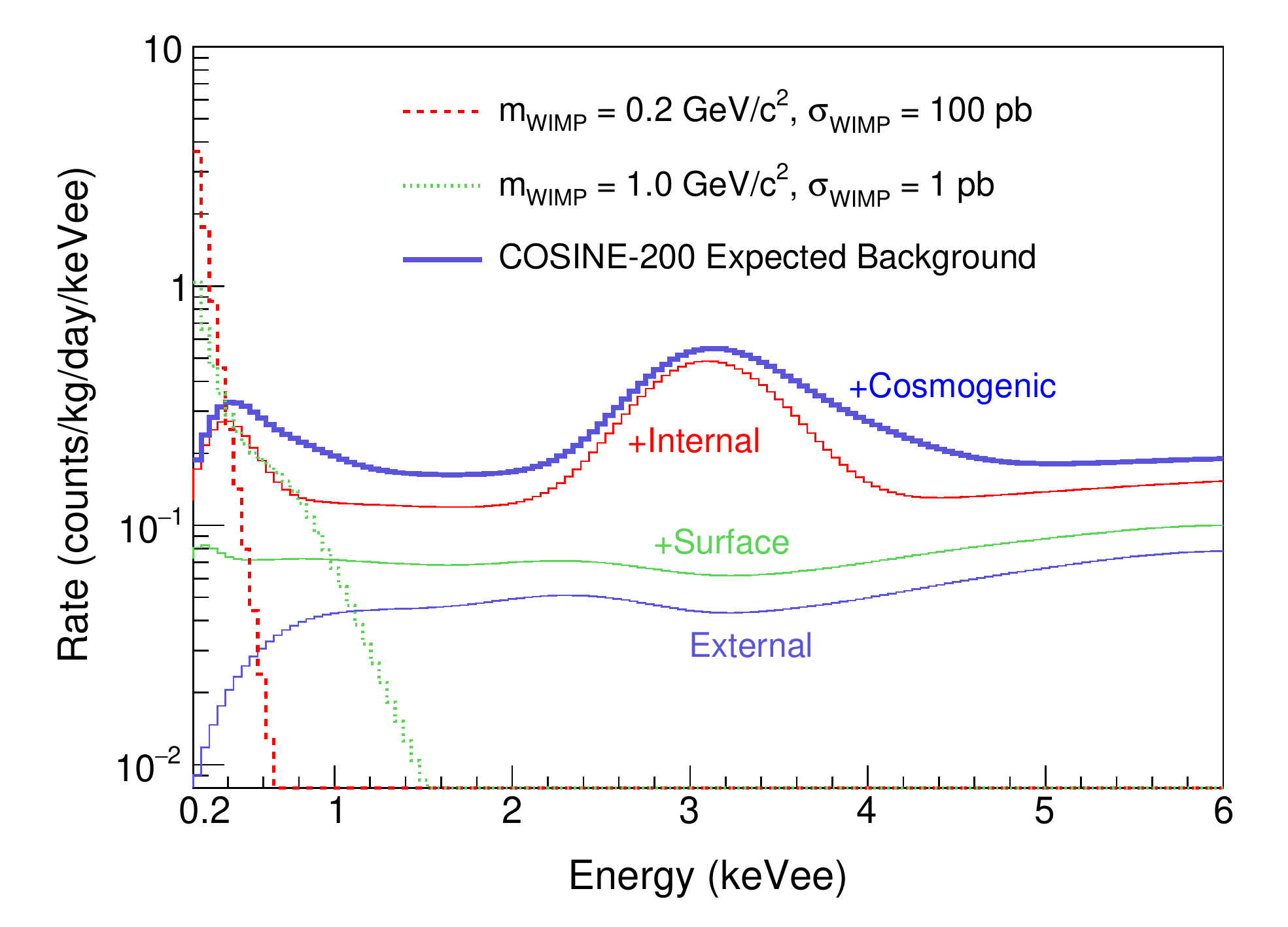} 
\caption{Expected background spectrum (blue solid line) of the COSINE-200 crystal based on the developed low-background NaI(Tl) crystal~\cite{COSINE:2020egt} is compared with expected signals using the Migdal effect for WIMP-nucleon SI interactions of  masses and cross sections of 0.2\,GeV/c$^2$, 100\,pb (red dashed line) and 1.0\,GeV/c$^2$, 1\,pb (green dotted line), respectively. A dominant background contribution is expected from the internal $^{40}$K and $^{210}$Pb remaining after purification. 
}
\label{fig_exp}
\end{center}
\end{figure}

An effort to upgrade the on-going COSINE-100 to the next-phase COSINE-200 has resulted in the production of NaI(Tl) crystals with reduced internal background from $^{40}$K and $^{210}$Pb~\cite{Shin:2018ioq,Shin:2020bdq,COSINE:2020egt} as well as an increased light yield of 22\,NPE/keV~\cite{Choi:2020qcj}. 
The recrystalization method has achieved chemical purification of the raw NaI powder with sufficient reduction of K and Pb contamination~\cite{Shin:2018ioq,Shin:2020bdq}. 
A dedicated Kyropoulos grower for small test crystals has produced low-background NaI(Tl) crystals with reduced $^{40}$K and $^{210}$Pb of less than 20\,ppb and 0.5\,mBq/kg, respectively, corresponding to background rates of less than 1\,counts/day/kg/keVee at the 1--6\,keVee ROI~\cite{COSINE:2020egt}. A full-size Kyropoulos grower has been built for the 100\,kg-size crystal ingot and will provide approximately 200\,kg of low-background NaI(Tl) detectors for the COSINE-200 experiment. The expected background level of those crystals is less than 0.5\,counts/day/kg/keVee in the ROI as shown in Fig.~\ref{fig_exp}. This estimate is based on the measured background levels for small test crystals as discussed in Ref.~\cite{COSINE:2020egt}.

A high light yield of the NaI(Tl) crystal is crucial for enabling low-energy thresholds below 1\,keVee. With an optimized concentration of thallium doping in the crystal, we achieved high light yield of 17.1$\pm$0.5\,NPE/keVee, slightly larger than that of the COSINE-100 crystal of approximately 15\,NPE/keVee. Further increase in the light collection efficiency by $\sim$50\% is possible with an improved encapsulation scheme as described  in Ref.~\cite{Choi:2020qcj} that directly connects the crystal and PMTs without an intermediate quartz window for which a 22\,NPE/keVee light yield is reported. 

The typical trigger requirement of the COSINE-100 experiment is satisfied with coincident photoelectrons in two PMTs attached to each side of the crystal at approximately 0.13\,keVee. 
However, the PMT-induced noise events are dominantly triggered below energies of a few keVee. 
The multivariable BDT provided a 1\,keVee analysis threshold with less than 0.1\% noise contamination and above 80\% selection efficiency~\cite{Adhikari:2020xxj}. 
A key variable in the BDT is the likelihood parameter using the event shapes of the scintillation-like events and the PMT-induced noise-like events. 
A further improvement of the low-energy event selection is ongoing with the COSINE-100 data by developing new parameters for the BDT as well as employing a machine learning technique that uses raw waveforms directly. 
COSINE-200 targets an analysis threshold of 5\,NPE~(0.2\,keVee), which is similar to the energy threshold that has already been achieved by the COHERENT experiment with CsI(Na) crystals~\cite{COHERENT:2017ipa}. 

The COSINE-200 experiment can be realized in a 4$\times$4 array of 12.5\,kg NaI(Tl) modules by replacing crystals inside the COSINE-100 shield~\cite{Adhikari:2017esn}. The COSINE-200 experiment will run at least 3 years for an unambiguous test of the DAMA/LIBRA annual modulation signals~\cite{Adhikari:2015rba}. In addition, this detector  can be used for general dark matter searches, especially for low-mass WIMPs via the Migdal process. Assuming the background reduction shown in Fig.~\ref{fig_exp}, a high light yield of 22\,NPE/keVee, and an analysis threshold of 5\,NPE corresponding to 0.2\,keVee, the sensitivities of the COSINE-200 experiment are evaluated below using the Migdal effect.

\begin{figure}[!htb]
\begin{center}
\includegraphics[width=0.49\textwidth]{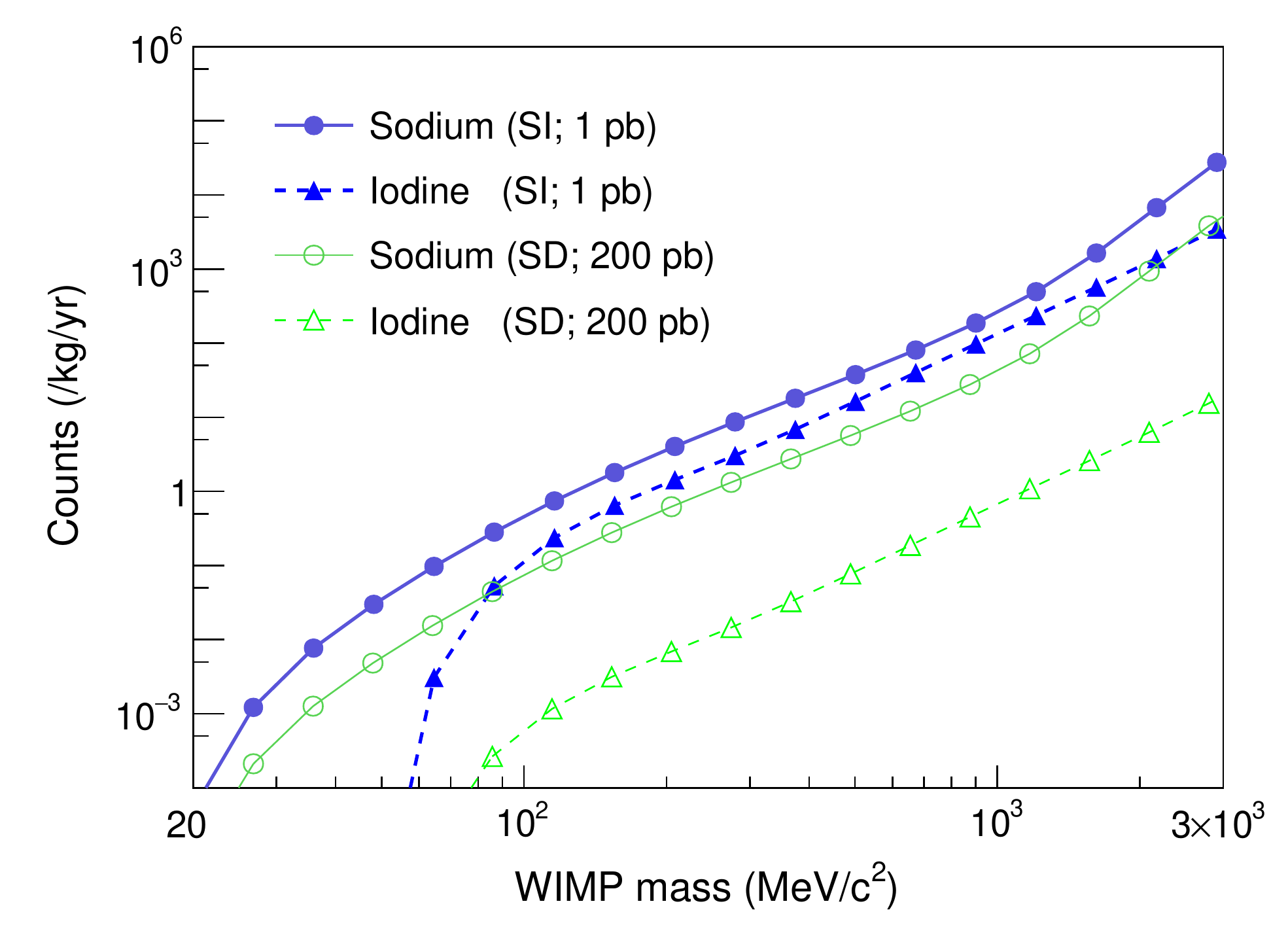} 
\caption{The expected signal counts using the Migdal effect for a one year of data assuming 0.2\,keVee analysis threshold are presented as a function of WIMP masses for SI (blue filled points) and SD interactions (green open points) for 1\,kg of sodium (circle points) and 1\,kg of iodine (triangle points). The cross-sections for the SI and SD interactions are assumed to be 1\,pb and 200\,pb, respectively.}
\label{fig_counts}
\end{center}
\end{figure}

\begin{figure*}[!htb]
\begin{center}
\includegraphics[width=0.98\textwidth]{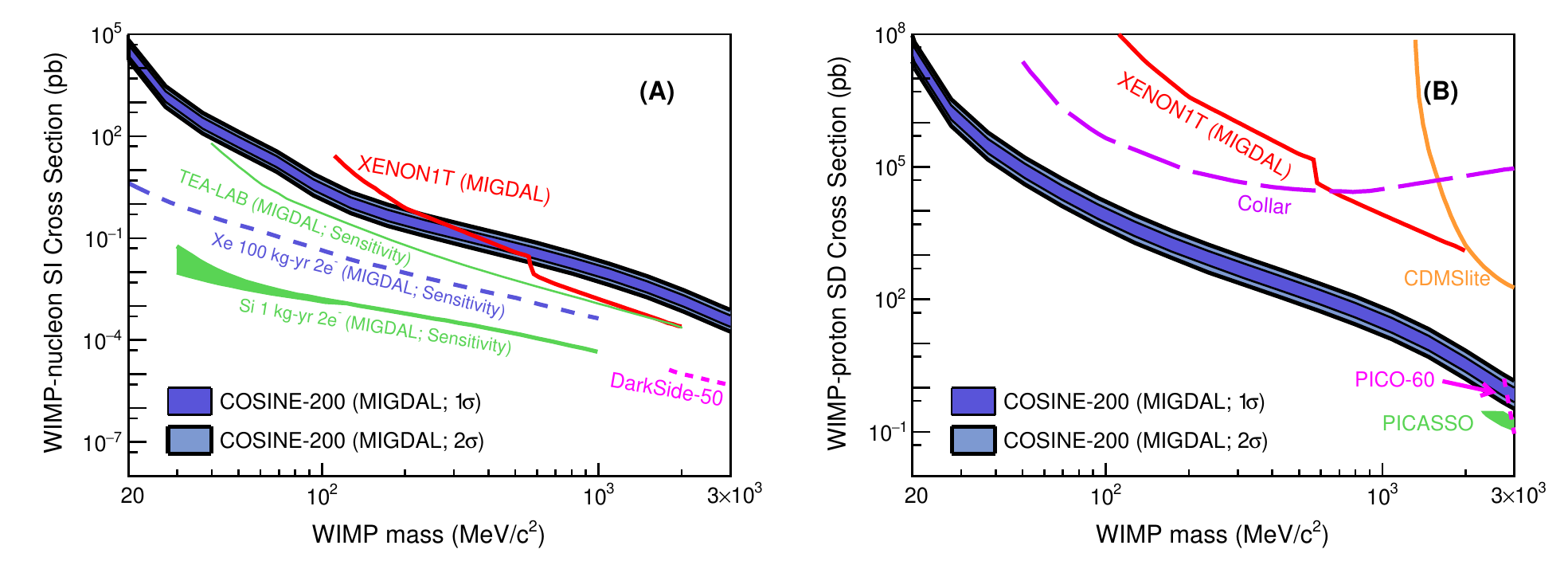}
\caption{COSINE-200 expected 90\% CL limits using the Migdal effect on the WIMP-nucleon SI cross-section (A) and the WIMP-proton SD cross-section (B) are presented assuming the background-only hypothesis indicating the 1$\sigma$ and 2$\sigma$ standard deviation probability regions over which the limits have fluctuated. Those limits are compared with the current best limits from the XENON1T Migdal~\cite{XENON:2019zpr}, DarkSide-50~\cite{Agnes:2018ves}, PICASSO~\cite{Behnke:2016lsk}, CDMSlite~\cite{SuperCDMS:2017nns}, Collar~\cite{Collar:2018ydf}, and PICO-60~\cite{PICO:2019vsc} together with the sensitivity limits from future liquid Ar experiment (TEA-LAB)~\cite{GrillidiCortona:2020owp}, Xe~\cite{Essig:2019xkx}, and Si~\cite{Knapen:2020aky} with Migdal.}
\label{fig_sensitivity}
\end{center}
\end{figure*}

We generate the WIMP interaction signals including the Migdal effects as discussed above. Here we assume a one year data exposure with 200\,kg crystals and the aforementioned detector performances. Poisson fluctuations of the measured NPE are considered for the detector resolution. Figure~\ref{fig_counts} presents the expected signal rate as a function of the WIMP mass for 1\,kg of  sodium and 1\,kg of iodine for different interactions assuming 0.2\,keVee threshold.  As one can see, sodium has an advantage for the low-mass WIMP searches with the Migdal process. 
In this scenario, the COSINE-200 experiment can probe the sub-GeV/c$^2$ WIMP with mass down to 20\,MeV/c$^2$.

We use an ensemble of simulated experiments to estimate the sensitivity of the COSINE-200 experiment, expressed as the expected cross-section limits for the WIMP-nucleon SI and WIMP-proton SD interactions using the Migdal effect in the case of no signals. For each experiment, we determine a simulated spectrum for a background-only hypothesis with assumed background from Fig.~\ref{fig_exp}. A gaussian fluctuation of each background component and a poisson fluctuation of each energy bin are considered for each simulated experiment. We then fit the simulated data with a signal-plus-background hypothesis with flat priors for the signal and Gaussian constraints for the backgrounds based on understanding of the NaI(Tl) crystals~\cite{cosinebg2,COSINE:2020egt}. Examples of signal spectra using the Migdal effect for SI interactions with WIMP masses (cross sections) of 0.2\,GeV/c$^2$ (100\,pb) and 1.0\,GeV/c$^2$ (1\,pb) are presented in Fig.~\ref{fig_exp}. The same Bayesian approach is used for the single-hit energy spectra between 5\,NPE (0.2\,keVee) and 130\,NPE (6\,keVee) for each WIMP model for several masses. The 1$\sigma$ and 2$\sigma$ standard-deviation probability regions of the expected 90\% CL limits are calculated from 2000 simulated experiments. Figure~\ref{fig_sensitivity} shows those 1$\sigma$ and 2$\sigma$ regions, for the COSINE-200 experiment using the Migdal effect. The limits on the WIMP-nucleon SI interaction shown in Fig.~\ref{fig_sensitivity} (A)  are compared with the current best limit on the low-mass WIMP searches of XENON1T with Migdal, DarkSide-50, and the expected sensitivities from future liquid Ar (TEA-LAB)~\cite{GrillidiCortona:2020owp}, Xe~\cite{Essig:2019xkx}, and Si~\cite{Knapen:2020aky} with Migdal.
We also verify no significant impact on the upper limits using the {\sc verne} code~\cite{verne}.
The COSINE-200 experiment has a potential to probe unexplored cross section values for the WIMP mass below 200\,MeV/c$^2$. 
In case of the WIMP-proton SD interactions shown in Fig.~\ref{fig_sensitivity} (B), our projected sensitivities are compared with the current best limits from XENON1T with Migdal, PICO-60, CDMSlite~\cite{SuperCDMS:2017nns}, Collar~\cite{Collar:2018ydf}, and PICASSO. Taking advantages of odd-proton numbers in both iodine and sodium, the projected sensitivities from the COSINE-200 experiments can probe unexplored parameter spaces of WIMP masses below 2\,GeV/c$^2$. 

\section{Conclusion}
We consider the Migdal effect to search for the low-mass dark matter using the COSINE-100 data. With 1\,keVee analysis threshold, our search extends down to 200\,MeV/c$^2$ WIMP mass. We have investigated the expected sensitivities of the COSINE-200 experiment with a total mass of 200\,kg, a 1-year period of stable operation, about 0.5 counts/day/keVee/kg background rate, and 0.2\,keVee energy threshold. In this scenario, the COSINE-200 detector can explore low-mass dark matter down to 20\,MeV/c$^2$, with a potential to probe currently unexplored parameter spaces for both SI and SD interactions.

\acknowledgments
We thank the Korea Hydro and Nuclear Power (KHNP) Company for providing underground laboratory space at Yangyang. This work is supported by: the Institute for Basic Science (IBS) under project code IBS-R016-A1 and NRF-2021R1A2C3010989, Republic of Korea; NSF Grants No. PHY-1913742, DGE1122492, WIPAC, the Wisconsin Alumni Research Foundation, United States; STFC Grant ST/N000277/1 and ST/K001337/1, United Kingdom; Grant No.  2017/02952-0 FAPESP, CAPES Finance Code 001, CNPq 131152/2020-3, Brazil.
\bibliographystyle{PRTitle}
\providecommand{\href}[2]{#2}\begingroup\raggedright\endgroup
\end{document}